\documentclass[journal=jctce,manuscript=article]{achemso}
\usepackage[version=3]{mhchem}

\usepackage{graphicx} 
\usepackage{subfig}
\usepackage{multirow}
\usepackage{longtable}
\usepackage{achemso} 
\usepackage{array,booktabs}
\usepackage{float}
\newcolumntype{C}[1]{>{\centering\arraybackslash}p{#1}}
\usepackage{caption}
\usepackage{python}
\usepackage{color}
\usepackage[]{siunitx}
\usepackage{enumitem} 
\usepackage{setspace}
\usepackage{verbatim} 
\setkeys{acs}{usetitle = true}
\usepackage{color, colortbl}
\definecolor{Gray}{gray}{0.9}
\usepackage{comment}
\usepackage{chemfig}
\usepackage{multirow}
\usepackage{multicol}
\usepackage[absolute,overlay]{textpos}
\usepackage{caption}
\usepackage{amsmath}
\usepackage{amsfonts}
\usepackage{amssymb}
\usepackage{graphicx}
\usepackage{algorithm}
\usepackage{algpseudocode}

\author{Vansh Ramani}
\affiliation{Department of Chemical Engineering, Indian Institute of Technology, Delhi,\\ Hauz Khas, New Delhi 110016, India}
\author{Tarak Karmakar}
\affiliation{Department of Chemistry, Indian Institute of Technology, Delhi,\\ Hauz Khas, New Delhi 110016, India}
\email{tkarmakar@chemistry.iitd.ac.in}

\title[\texttt{achemso}]
{Graph Neural Networks for Predicting Solubility in Diverse Solvents using MolMerger incorporating Solute-solvent Interactions}

\begin{document}
\newpage
\begin{abstract}
Prediction of solubility has been a complex and challenging physiochemical problem that has tremendous implications in the chemical and pharmaceutical industry. Recent advancements in machine learning methods have provided great scope for predicting the reliable solubility of a large number of molecular systems. However, most of these methods rely on using physical properties obtained from experiments and or expensive quantum chemical calculations. Here, we developed a method that utilizes a graphical representation of solute-solvent interactions using `MolMerger', which captures the strongest polar interactions between molecules using Gasteiger charges and creates a graph incorporating the true nature of the system. Using these graphs as input, a neural network learns the correlation between the structural properties of a molecule in the form of node embedding and its physiochemical properties as output. This approach has been used to calculate molecular solubility by predicting the Log solubility values of various organic molecules and pharmaceuticals in diverse sets of solvents.  
    
\end{abstract}

\newpage
\section{INTRODUCTION}
The molecular solubility of a solute is an important property that plays a crucial role in material science, environmental chemistry, food and beverage industry, chemical process optimization, biotechnology, cosmetic formulation, agrochemicals, oil and gas Industry, and drug development. The prediction of molecular solubility of a solute in various solvents is one of the crucial challenges that attracted a lot of attention from the research community.  A robust solubility model precisely forecasting solubility issues during the early stage of drug development process significantly reduces time and monetary resources. A trustworthy solubility model aids in the selection of promising drug candidates during the drug discovery process, halting the advancement of compounds with low solubility that could cause problems with formulation and or lower bio-availability. Understanding solubility patterns helps chemists create formulations with the best possible drug dissolution rates, which is essential for both drug efficacy and patient outcomes.

Solubility prediction has been a long-standing problem\cite{llinas2008solubility,hewitt2009silico,llinas2020findings}, and over the years a plethora of methods have been developed to predict solubility of molecular systems.\cite{hildebrand1950solubility,klamt2002prediction,sanghvi2003estimation,hansen2007hansen,delaney2005predicting,dearden2006silico,louis2010prediction,gupta2011prediction,liu2016using,li2018computational,alsenz2019quantum,francoeu2021soltrannet,bjelobrk2021solubility,bjelobrk2022solubility,reinhardt2023streamlined} Among the recent ML-based methods, Graph Neural Networks (GNNs), including Graph Convolution Networks (GCN), Message Passing Neural Networks (MPNN), Graph Attention Networks (GATs), and Attentive Finger Printing (AFP) have gained enormous popularity due to their inherent capability of embedding structural properties of molecules as graphs.\cite{lusci2013deep,coley2017convolutional,withnall2020building,cui2020improved,tang2020self,gao2020accurate,boobie2020machine,lee2022novel,lee2023multi,panapitiya2022evaluation,ahmad2023attention} These methods mostly utilize structural as well as experimentally obtained physicochemical properties to predict the solubility of molecular systems. Additionally, they are limited to predicting aqueous solubility except in a handful of cases where non-aqueous solvents have been considered. In a recently proposed method, Lee \textit{et al.} incorporated solute-solvent pairs in a GCN, which passes both the solute and solvent information through convolution layers and passes them through a Multi-Layer Perceptron (MLP) by concatenating them into a single one-dimensional embedding, as input neurons.\cite{lee2022novel} The problem with this approach is that, with a complex neural network, the model categorizes solubilities by solvent and recognizes patterns using the solvents in the training set. When a new solvent is introduced, the model fails, due to its over-fitting to known solvents. Thus there has been a need for developing a new method that can incorporate physical interactions between solute and solvent without overfitting or categorizing solubility to specific solvents, and while message passing, the interactions between the solute and solvent, such as polar, hydrogen, or ionic interactions can be included in the input graph.

In this work, we have developed a GNN-based model that takes into account explicitly the solute-solvent interactions and provides reliable solubility of a large number of organic molecules including active pharmaceuticals in both aqueous and organic solvents. The solute-solvent interactions have been incorporated by a graph `MolMerger', which by using Gastieger charges\cite{gasteiger1980iterative} incorporates interactions between two polar, partially charged atomic centers in the solute and solvent molecules. This way the GNN learns the correlation between the structures of solute and solvent molecules as well as their physical interactions, thereby providing accurate solubility values for a solute in several solvents.

\section{METHODS}

\subsection{Training Data}
We have collated the dataset (total size of 5198) from three different sources - BigSolDB\cite{krasnov-2023}, BNNLabs Solubility, and ESOL\cite{10.1021/ci034243x}. Our aim here is to keep a limited set of solvents while training and evaluate a larger set of solvents to see the efficiency of the model, and if the model can predict never-seen-before solvents. The most number of data entries in Table 1 are from BigSolDB. The 13 solvents each having solutes ranging from 127 to 1212. The descriptor used is the canonical SMILES representation. Each solute-solvent pair was converted to a Molecular Fingerprint using RDKit Library in Python.

\begin{figure}
    \centering
    \includegraphics[width=0.7\linewidth]{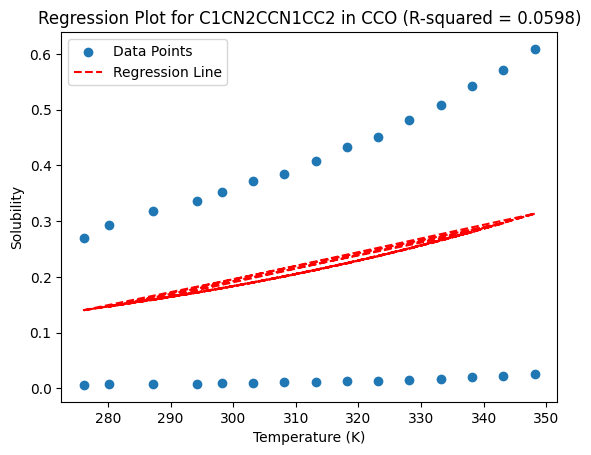}
    \caption{Examples of erroneous data points in the dataset.}
    \label{fig:badata}
\end{figure}

The data collected from BigSolDB needed to be cleaned to avoid false and bad data from the database. The BigSolDB dataset has solubilities for a solute-solvent pair at 1 to 15 different temperatures. The solute-solvent pairs were grouped. For each pair, solubilities were plotted against temperature. On inspection, it was observed that a large set of solute-solvent pairs had random data points with redundant values as shown in Figure \ref{fig:badata}. Some solvents had more than one LogS value for each solute at different temperatures. Since a manual inspection is not possible for a dataset of 50000 data points, a robust non-biased algorithm was applied. A regression was fit to each solute-solvent pair, with a general formula:
\vspace{-5mm}
\begin{equation}
    S = ae^{bT}
\end{equation}
where $S$ is solubility, and $a$ and $b$ are parameters discussed in article\cite{doi:10.1021/op100006y}.

Solute-solvent pairs with $R^2$ accuracy of less than 0.925 were removed, constituting less than 2\% of the data. These 2\% plots were manually inspected to ensure no loss of valuable data takes place. Further, both solute and solvent were converted to an RDKit molecule using the `MolFromSmiles' method, and SMILES with errors were removed. SMILES with more than one molecule represented by a "." were also discarded. Solvents with a solute frequency greater than 120 were used for training (split into the train and test sets), and less than 120 and greater than 2 were kept for a robust evaluation on a broader set. Finally, the LogS value of a solute-solvent pair, at a temperature closest to 273K was used. It is important to note that, even optimization was done using metrics on the test set isolated from the training set, which ensures the evaluation data is untouched.

\begin{figure}[!ht]
    \centering
    \includegraphics[width=0.7\linewidth]{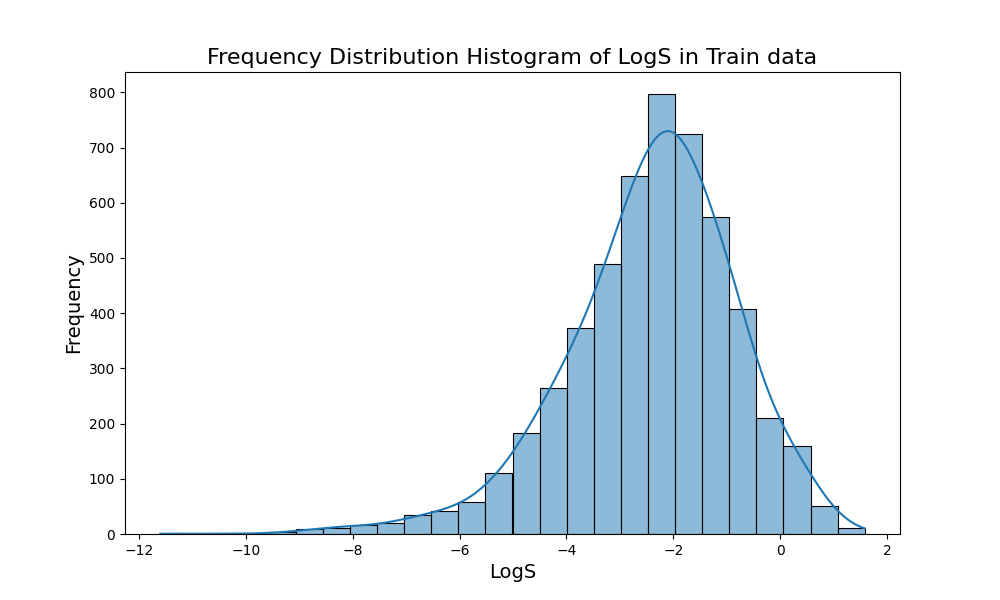}
    \includegraphics[width=0.7\linewidth]{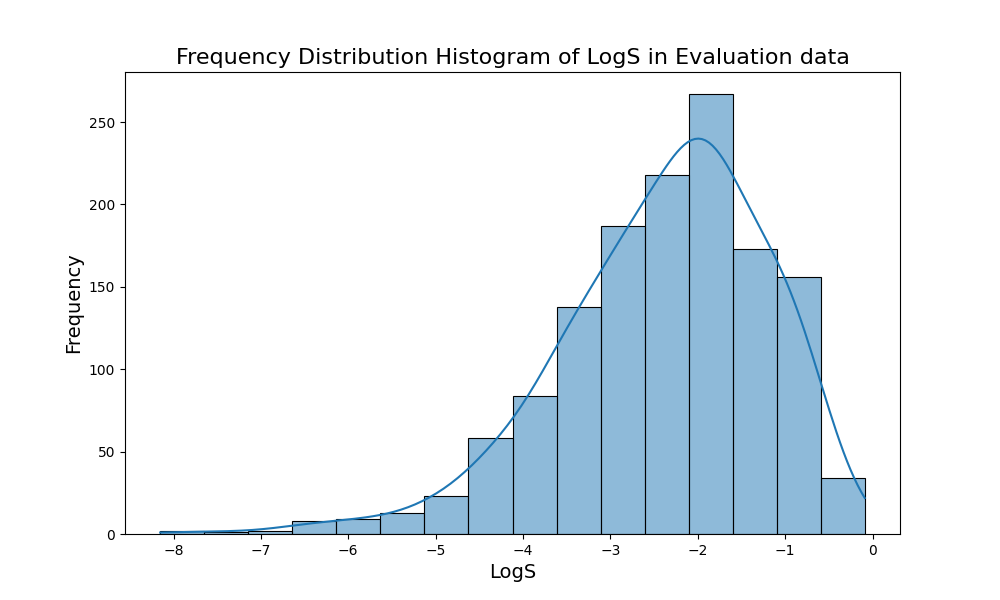}
    \caption{The distributions of the LogS values in the (a) training and (b) evaluation datasets.}
    \label{fig:LogS-distr}
\end{figure}

The distributions of the LogS values in the training and the evaluation datasets are shown in Figure \ref{fig:LogS-distr}. The most probable values of LogS range from -4.5 to 0 in both the training and evaluation datasets.

\vspace{5mm}
  \begin{table}[h]
  \centering
  \begin{tabular}{|c|c|c|c|c|}
  \hline
  Serial No. & Source & No. of Solvents & No. of Data Entries & \% of total data \\
 \hline
  1 & BigSolDB & 10 & 3126 & 60.1\% \\
  2 & BNNLabs Solubility & 2 & 1093 & 21.0\% \\
  3 & ESOL & 1 & 979 &  18.9\%\\

    \hline
    \end{tabular}
    \caption{Sources for Data}
    \label{tab:mytable}
    \end{table}

    \vspace{4mm}
    \begin{table}[h]
    \centering
    \begin{tabular}{|c|c|c|c|}
    \hline
    Serial No. & Solvent SMILES & Solvent Name & Solute Frequency \\
    \hline
    1 & C1COCCO1 & Tetrahydrofuran & 127 \\ 
    2 & CC\#N & Acetonitrile & 237 \\ 
    3 & CC(=O)C & Acetone & 427 \\ 
    4 & CC(C)=O & Acetaldehyde & 299 \\ 
    5 & CC(C)O & Ethanol & 329 \\ 
    6 & CCCCCO & Hexane & 255 \\ 
    7 & CCCO & Propanone & 274 \\ 
    8 & CCO & Ethylene glycol & 1079 \\ 
    9 & CCOC(C)=O & Ethyl acetate & 331 \\ 
    10 & CN(C)C=O & DMF & 128 \\ 
    11 & CO & Carbon monoxide & 366 \\ 
    12 & Cc1ccccc1 & Toluene & 134 \\ 
    13 & O & Water & 1212 \\ 
    \hline
    \end{tabular}
    \caption{Solvents and Solute Frequency in Train Data
}
    \label{tab:mytable}
    \end{table}


\newpage
\subsection{Features of Graph and Graph Representation}
Graph Neural networks (GNNs) work in a way that makes it convenient to feed graph data to a neural network. A graph in Graph Neural Networks (GNNs) is represented as \( G = (V, E) \), where $V$ is the set of nodes (vertices) representing entities or elements in the graph, and $E$ is the set of edges representing relationships or connections between nodes. In the case of molecular systems, the atoms are described as nodes while the bonds are represented by edges. Graph representations not only make it easier to describe molecules, but attentive weights assigned to edges after message passing, help find the important regions in the graph by assigning higher weights, whereas less important sections of the graph with lower weights. In our case, only graph-level features and no global embedding have been used, which helps in finding the correlation between just the structural properties of a molecule and its property - the solubility. 

\begin{table}[h]
    \centering
    \setlength{\extrarowheight}{3pt}
    \begin{tabular}{|c|c|c|p{8cm}|}
    \hline
    Serial No. & Feature Name & Feature Size & Feature Description \\
    \hline
    1) & Type of Atom & 10 & \small{A vector of 1s and 0s for atom type. Set of Atoms = [“C”, “N”, “O”, “F”, “P”, “S”, “Cl”, “Br”, “I”, Unknown]}  \\
    2) & Hybridisation & 3 & {A vector of 1s and 0s for hydridisations. Set of Hybridisations = ["SP1", "SP2". "SP3"]} \\
    
    3) & Formal Charge & 1 & \small{A vector of a float value of formal charge. Set of attribute = ["Formal Charge"]} \\
    
    4) & Acceptor/Donor & 2 & \small{A vector of 1s and 0s of electronic behavior. Set of attributes = ["Acceptor", "Donor"]} \\
    
    5) & In Aromatic & 1 & \small{A vector of 1s and 0s for aromaticity. Set of attribute = ["Is in Aromatic System"]} \\
    
    6) & Degree & 7 & \small{A vector of 1s and 0s for degree of atom. Set of Degrees = ["0", '1", "2", "3", "4", "5", Unknown]} \\
    
    7) & Chirality & 2 & \small{A vector of 1s and 0s for chiral behavior. Set of Chiralities = ["R", "S"]} \\
    
    8) & No of Hydrogens & 5 & \small{A vector of 1s and 0s for hydrogens on atom. Set of count = ["0", "1", "2", "3", "4"]} \\
    \hline    
    \end{tabular}
    \caption{Node Features of Graph- Properties of atom}
    \label{tab:nodef}
\end{table}

\begin{table}[h]

    \setlength{\extrarowheight}{2pt}
    \resizebox{\textwidth}{!}{
    \begin{tabular}{|c|c|c|p{8cm}|}
    \hline
    Serial No. & Feature Name & Feature Size & Feature Description \\
    \hline
    1) & Type of Bond & 5 & \small{A vector of 1s and 0s for bond type. Set of bonds = [“SINGLE”, “DOUBLE”, “TRIPLE”, “AROMATIC”, "HYDROGEN" or Unknown]}  \\
    2) & Is in same ring & 1 & \small{A vector of 1s and 0s for is in same ring as atom. Set of attributes = ["Is in the ring"]} \\
    
    3) & Is in Conjugation & 1 & \small{A vector of 1s and 0s for is in conjugation. Set of attribute = ["is in conjugation"]} \\
    
    4) & Stereo Configuration & 2 & \small{A vector of 1s and 0s for stereochemistry of bond. Set of attributes = [“STEREONONE”, “STEREOANY”, “STEREOZ”, “STEREOE”, Unknown]} \\

    5) & Graph Distance & 1 & \small{A vector of 1s and 0s for graph distance. Set of attribute = [“1", "2" ... "7", higher]} \\    
    \hline
    \end{tabular}
    }
    \caption{Edge Features of Graph- Properties of bond}
    \label{tab:edgef}
\end{table}

A featurizer has been used to transform the 2D molecular structure into an object representation suitable as an input for the GNN. The object has three attributes - (1) node features (see Table \ref{tab:nodef}), (2) edge features (Table \ref{tab:edgef}), and (3) edge weights in the form of an adjacency Matrix. This adjacency matrix is used as a mask that represents connected nodes in a Graph. Most research on physiochemical properties prediction using Machine Learning and Deep Learning methods such as in Ref.\citenum{boobier-2020}, obtained commendable results, however, they mostly rely on using experimental data such as zero-point energies, solvation energies, Gibbs free energies, dipole moments, and solvent accessible surface area as features. Though these features increase accuracy, they are obtained from expensive quantum chemical calculations or sophisticated experiments. In our approach, we rely on using only the structural information of a molecule along with a Site-Charge (SC) description (discussed in the subsequent section) with no trivial correlation to solubility that can be computed from basic chemistry, without quantum calculations, MD simulations, and or experiments.

\subsection{Solute-solvent interactions: Site-Charge (SC) model}
Incorporation of solvent details into the NN model is crucial, especially, when one wants to calculate the solubility of a solute in different solvents. In Ref.\citenum{lee2022novel}, Lee \textit{et al.} studied a combination of solute and solvents to predict solubility by using a multi-input network model with solute and solvent fingerprints (GCNs). They showed that in addition to using two GCNs for solute and solvent, the inclusion of physicochemical property-based features improves the solubility prediction. This model, however, relies on internally classifying the solute-solvent pairs based on solvent properties. While this provides satisfactory results in terms of predicting the solubility, the models encounter limitations when a completely new solvent is considered. One way of avoiding this issue is to have a model that can learn solute-solvent interactions and predict solubility without exclusively knowing the structure of the solvent molecule.

\begin{figure}[!ht]
    \centering
    \includegraphics[scale=0.264]{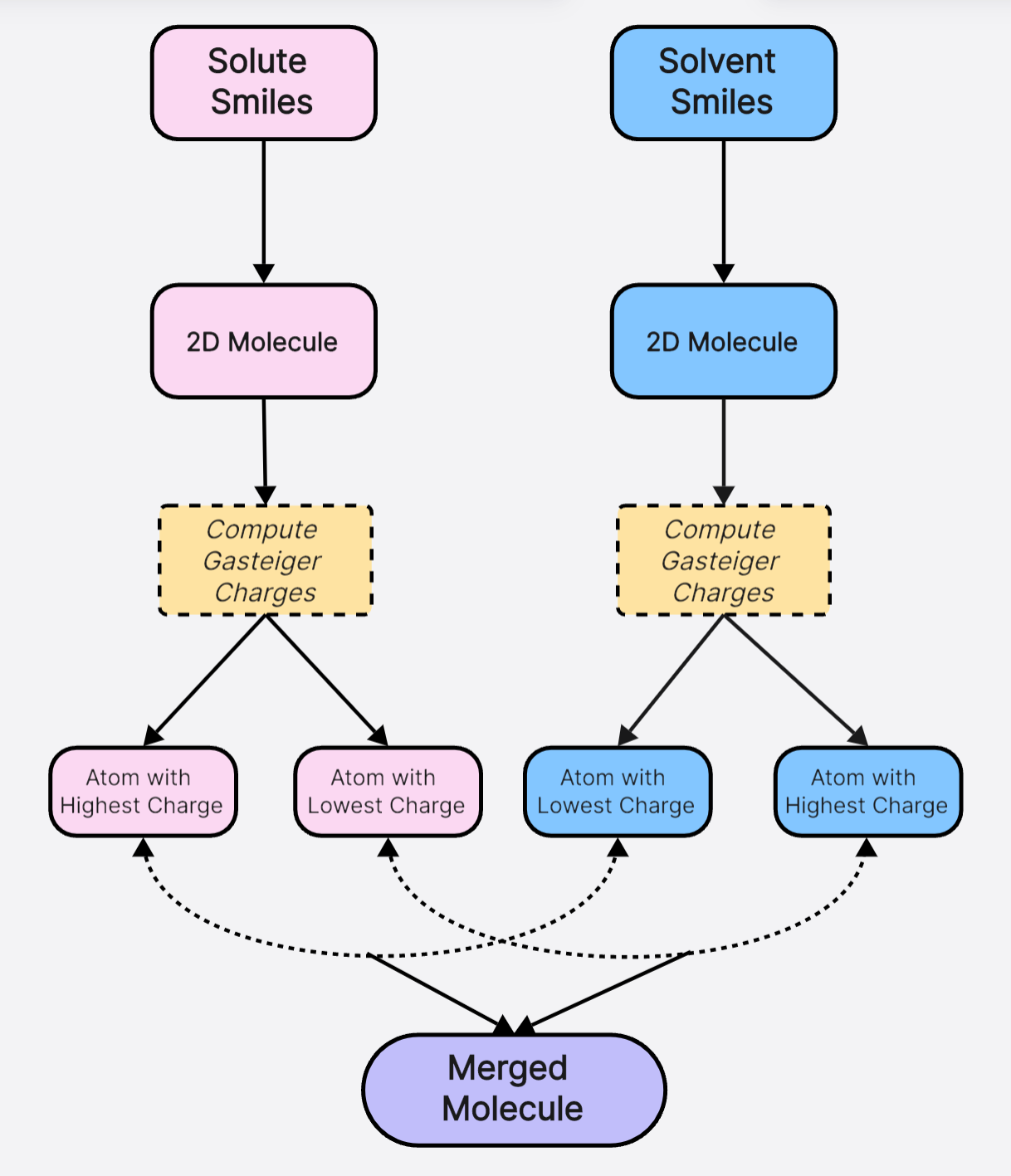}
    \includegraphics[scale=0.3]{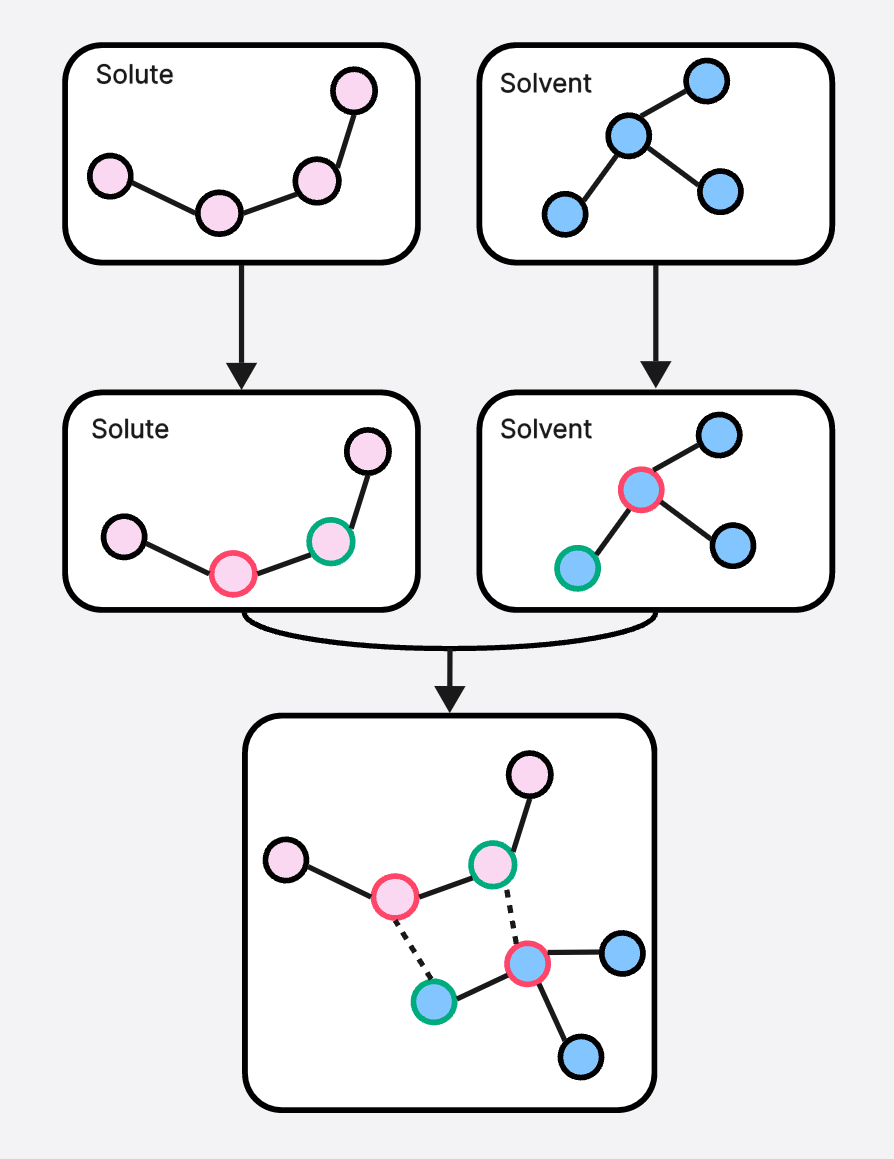}
    \caption{a) FlowChart of MolMerger Algorithm b) Depiction of working of MolMerger}
    \label{fig:MolMergFlow}
\end{figure}

Here, we have introduced a new way to incorporate solute-solvent interactions and the method we named `MolMerger'. The MolMerger algorithm takes in the RDKit 2D molecular representation of a solute and solvent and iterates over each atom in both molecules independently. (Figure \ref{fig:MolMergFlow}) During this iteration, the algorithm calculates the atomic partial charges using the Gastieger charges method as shown in Ref. \citenum{gasteiger1980iterative}. Table 5 shows the algorithm deployed for the calculation.  

\begin{figure}
    \centering
    \includegraphics[width=0.8\linewidth]{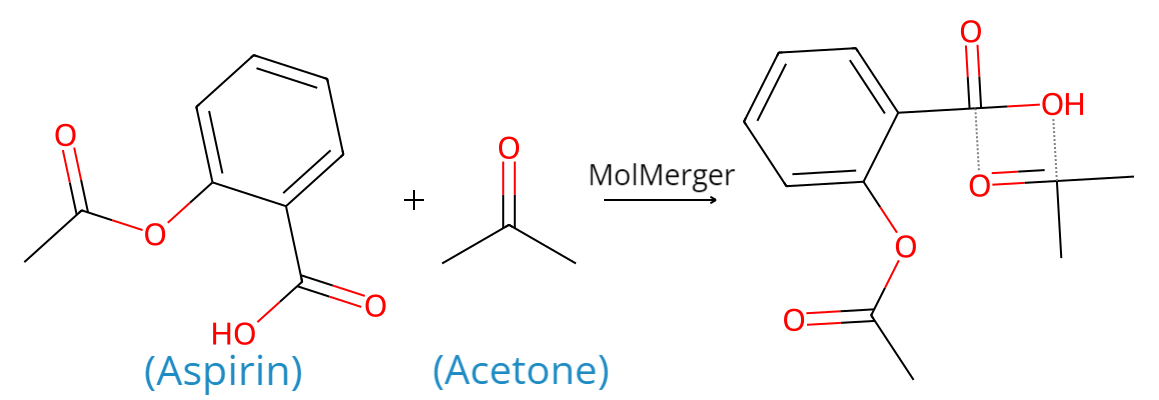}
    \caption{Example showing MolMerger working on Aspirin and Acetone.}
    \label{fig:eval}
\end{figure}

\begin{table}[h!]
\centering
\setlength{\extrarowheight}{0pt}
\begin{tabular}{|p{3cm}|p{7cm}|p{6cm}|}
\hline
\textbf{Step} & \textbf{Equation} & \textbf{Target} \\
\hline 
Initial charges: & $q_i = Z_i + \sum_{j \neq i} \dfrac{Z_j}{R_{ij}}$ & \footnotesize{Assigns seed charges based on atomic number and neighboring atoms. Initial charge $q_i$ for atom $i$.} \\
\small{Charge redistribution:} & $q_{i,k} = q_{i,k-1} + \alpha_i \left( \sum_{j \neq i} \dfrac{q_{j,k-1}}{R_{ij}^2} - \rho_i \right)$ $\alpha_i = \dfrac{1}{1 + \gamma_i \left| q_{i,k-1} \right|}$ & \footnotesize{Redistributes charges based on electronegativities and orbital configurations iteratively until convergence.} \\
&   & \\
Partial charges: & $q_i = Z_i + \sum_{j \neq i} q_j$ & \footnotesize{Final charges model electrostatic properties.} \\
\hline
\end{tabular}
\caption{Gasteiger Charges Method Steps}
\end{table}

\begin{table}[h!]
\centering
\begin{tabular}{|l|l|}
\hline
\textbf{Variable} & \textbf{Meaning} \\
\hline
$Z_i$ & Atomic number of atom $i$ \\
$R_{ij}$ & Distance between atoms $i$ and $j$ \\
$q_i$ & Charge of atom $i$ \\
$q_{i,k}$ & Charge of atom $i$ at iteration $k$ \\
$\alpha_i$ & Scaling factor for charge redistribution of atom $i$ \\
$\gamma_i$ & Damping factor for controlling convergence rate \\
$\rho_i$ & Electronegativity of atom $i$ \\
\hline
\end{tabular}
\caption{Variables and Meanings in Gasteiger Charge Method}
\end{table}

The atoms in solute and solvent with the highest and lowest Gasteiger charges are tagged, and a `\textit{virtual}' bond is formed between the most electron-dense atom in the solute to the least electron-dense atom in the solvent, and \textit{vice versa}. This approach captures and simplifies the complex polar interactions between solute-solvent molecules into two bonds in a new `\textit{Merged-Molecule}'. The two virtual bonds thus created between the solute and solvent make the Merged-Molecule unique for a specific solute-solvent pair.

\subsection{Model Description}
\begin{figure}
    \centering
    \includegraphics[width=1\linewidth]{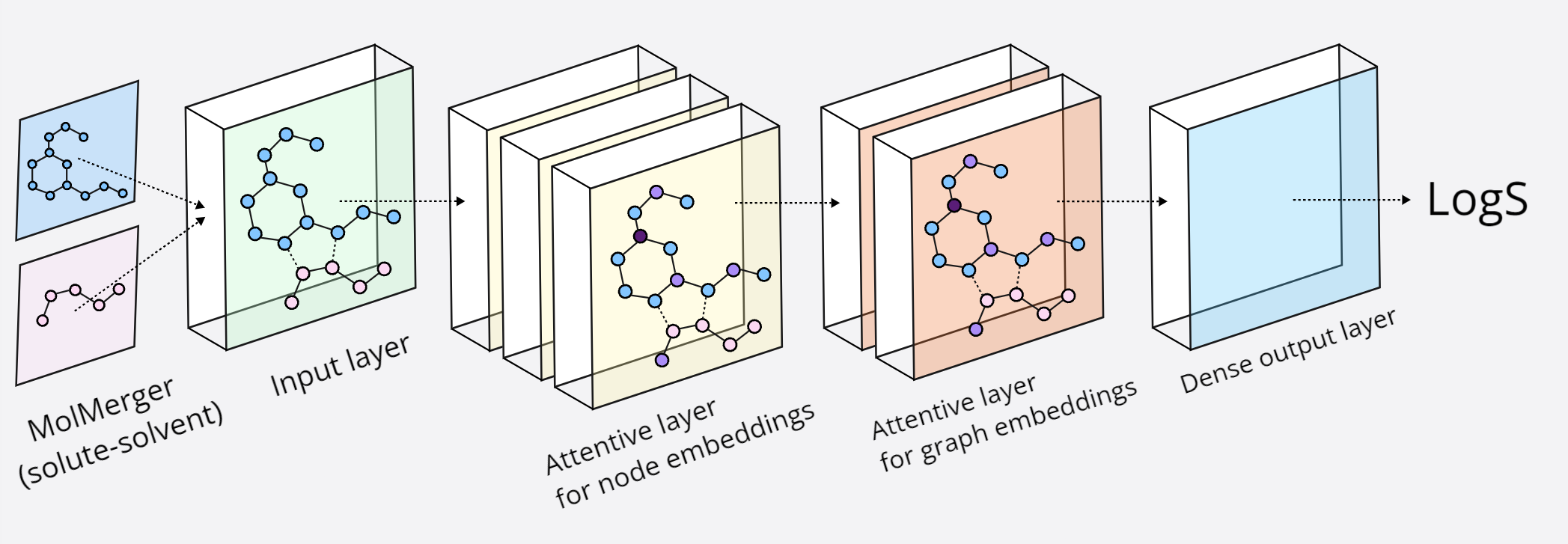}
    \caption{AttentiveFP model architecture}
    \label{fig:attentivefp}
\end{figure}

For the prediction of LogS after featurization on the merged molecule, the graph obtained is fed to a model with architecture based on the AttentiveFP model as described in Ref.\citenum{xiong2019}. The AttentiveFP model is a GNN model for molecular representation learning as depicted in Figure \ref{fig:attentivefp}. It is implemented in PyTorch Geometric and consists of the following components:
\begin{itemize}
    \vspace{-3mm}
    \item A linear layer that maps the input feature dimensionality to the hidden feature dimensionality.
    \vspace{-3mm}
    \item A GATEConv layer (gate-conv) that updates the node features based on the edge features and the previous node features.
    \vspace{-3mm}
    \item A GRU cell (gru) that updates the node features based on the output of the GATEConv layer.
    \vspace{-3mm}
    \item A list of GATConv layers (atom-convs) and GRU cells (atom-grus) that further update the node features.
    \vspace{-3mm}
    \item A GATConv layer (mol-conv) that generates a molecular embedding by pooling the updated node features.
    \vspace{-3mm}
    \item A GRU cell (mol-gru) that further updates the molecular embedding.
    \vspace{-3mm}
    \item A linear layer that maps the molecular embedding to the output feature dimensionality.
\end{itemize}

The model takes as input the node features (x), edge indices (edge-index), and edge features (edge-attr), as well as a batch vector (batch) that defines the molecular structure of the input data. It outputs a vector representation of each molecule. The forward method applies a series of graph convolution and GRU operations to the node and molecular features and uses dropout for regularization. The reset-parameters method resets the learnable parameters of the model to their initial values.

\subsection{Attentive Fingerprinting}
\begin{figure}
    \centering
    \includegraphics[width=0.43\linewidth]{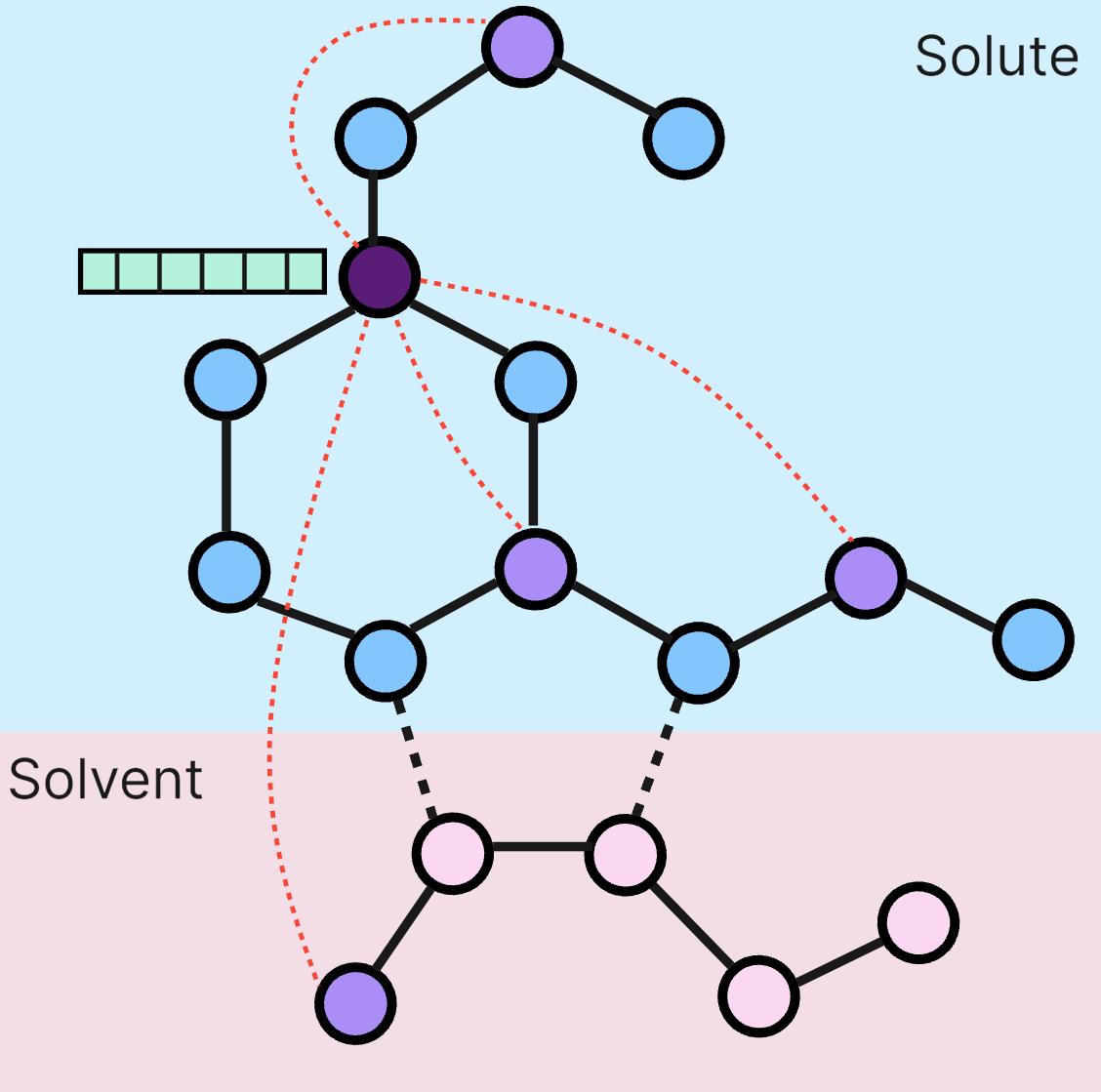}
    \includegraphics[width=0.23\linewidth]{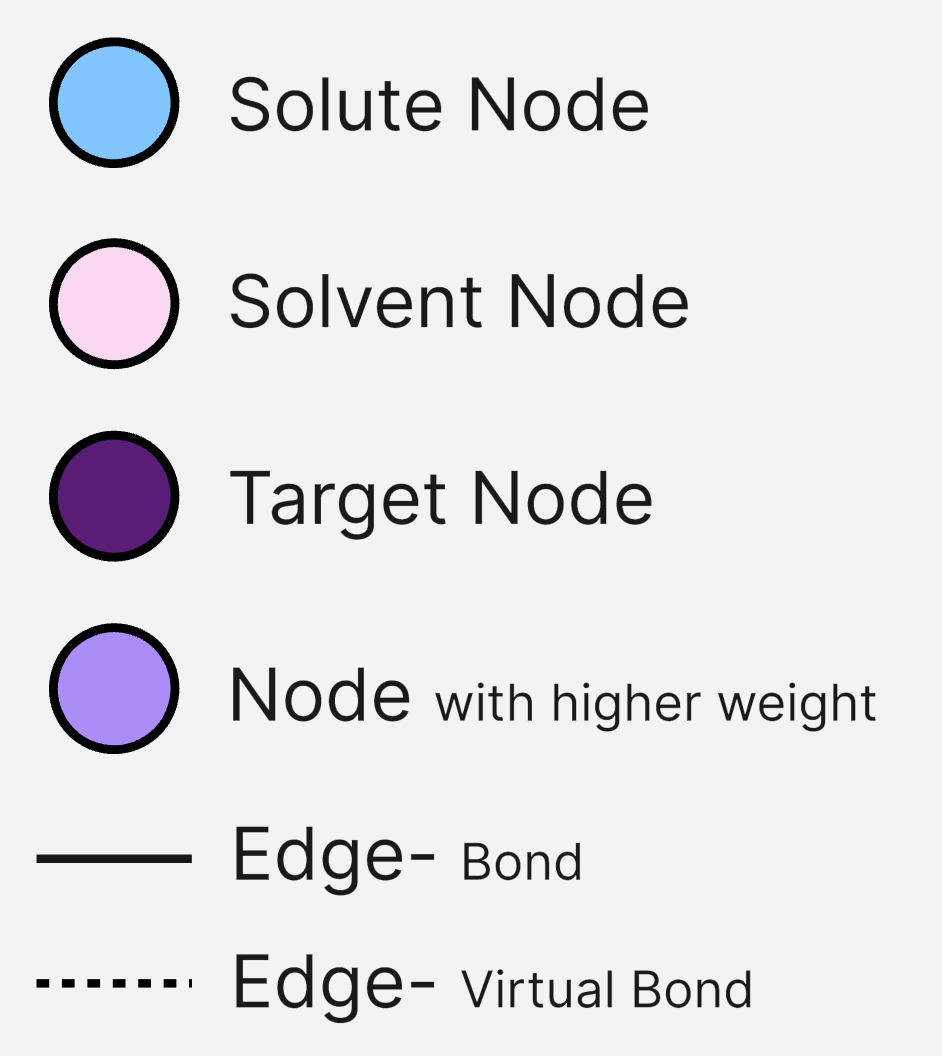}
    \caption{Depiction of attention context passing from nodes in solvent to target node in solute incorporating solute-solvent interactions}
    \label{fig:attentive}
\end{figure}

The AttentiveFP model\cite{xiong2019} is built on the Graph Attention Mechanism, which allows it to focus on the most relevant parts of the molecular graph when generating a molecular representation. By weighing the importance of different parts of the graph, the model can better capture the complex relationships between atoms and bonds in a molecule. (Figure \ref{fig:attentive}) This is in contrast to traditional GNNs, which treat all parts of the graph equally and may not be as effective at capturing the nuanced relationships in molecular graphs. The AttentiveFP uses gated recurrent units (GRUs) as it allows for the modeling of temporal dependencies in the graph-level molecular embedding. The GRU cell updates the molecular embedding based on the output of the GATConv layer, which captures the structural and chemical properties of the molecule at the atom level. By incorporating a GRU cell, AttentiveFP can capture how these atom-level properties evolve, which can be important for modeling certain molecular properties. In comparison to GAT, AttentiveFP may be better at capturing temporal dependencies in the data. \\

\newpage
\subsection{Loss Function and Hyperparameters}
For the regression task of LogS prediction, L2Loss was used. Whereas for the Classification task, Sparse Softmax CrossEntropy with Logits (SSCE) was used as described in equations 2 and 3:
\begin{align}
\text{L2 Loss:} & \quad L(\hat{y}, y) = \sum_{i=1}^{n} \frac{1}{2} \lVert \hat{y} - y \rVert^{2} 
\end{align}
\vspace{-15mm}
\begin{align}
\text{SSCE Loss:} & \quad L_{\text{SSCE}}(\mathbf{z}, \mathbf{y}) = -\sum_{i} y_i \log\left(\frac{e^{z_i}}{\sum_{j} e^{z_j}}\right) \\
\mathbf{z} &\text{ : input logits} \nonumber \\
\mathbf{y} &\text{ : sparse ground truth label vector.} \nonumber
\end{align}

The number of layers for Graph Attention, the number of time steps, and the number of epochs are the model's hyperparameters. The number of layers and time steps are important for the internal architecture of the model, which determines the complexity of the model. It is worth noting that a more complex model is not necessarily better, as a huge amount of data is required to fit the parameters. The number of epochs must be optimized based on the loss in the test data, to avoid over-fitting to the train data.

\section{RESULTS AND DISCUSSION}

\begin{figure}[!ht]
    \centering
    \includegraphics[width=0.6\linewidth]{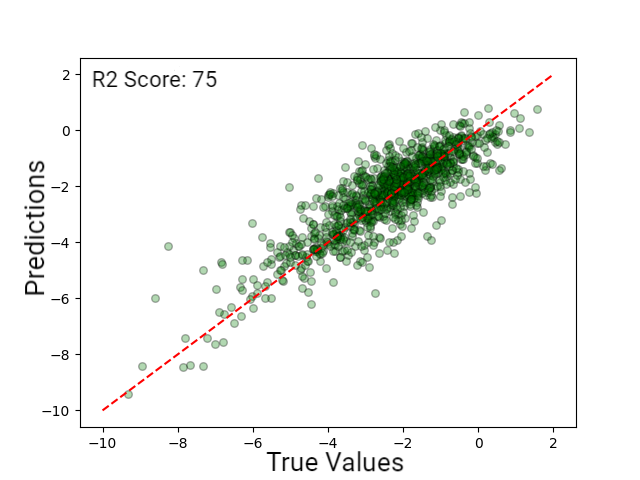}
    \caption{Predicted vs experimental LogS values on test set}
    \label{fig:regres}
\end{figure}

\begin{figure}
    \centering
    \includegraphics[width=0.47\linewidth]{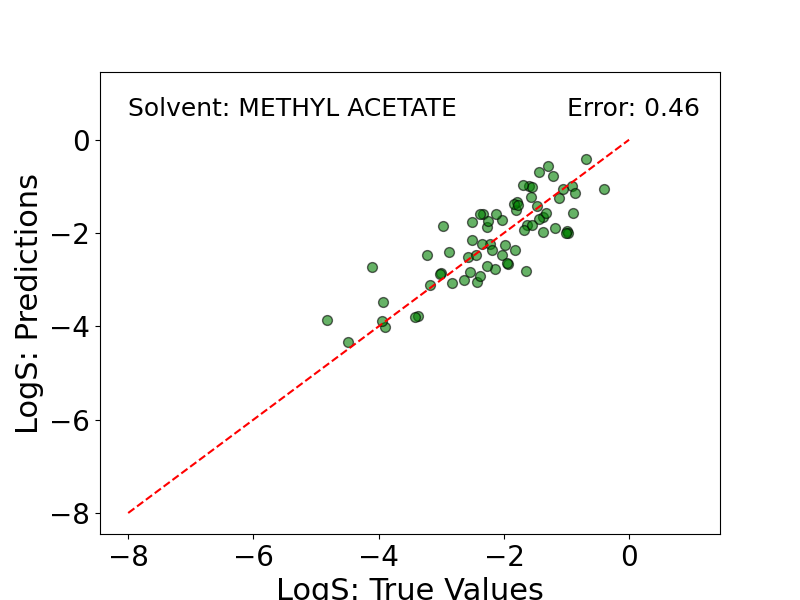}
    \includegraphics[width=0.47\linewidth]{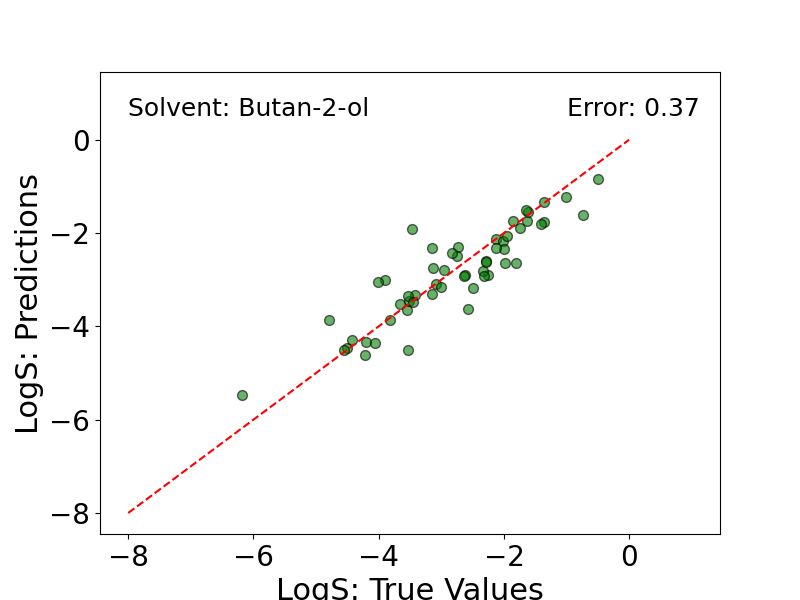}
    \includegraphics[width=0.46\linewidth]{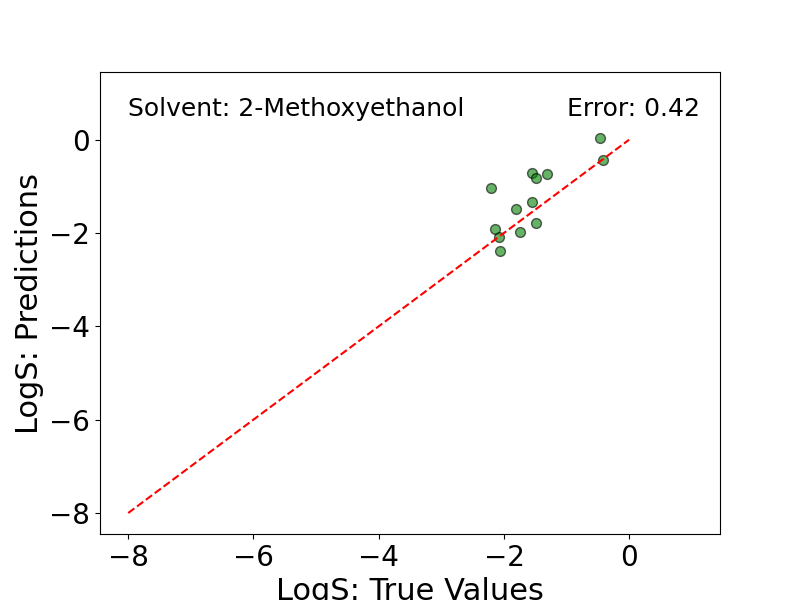}
    \includegraphics[width=0.46\linewidth]{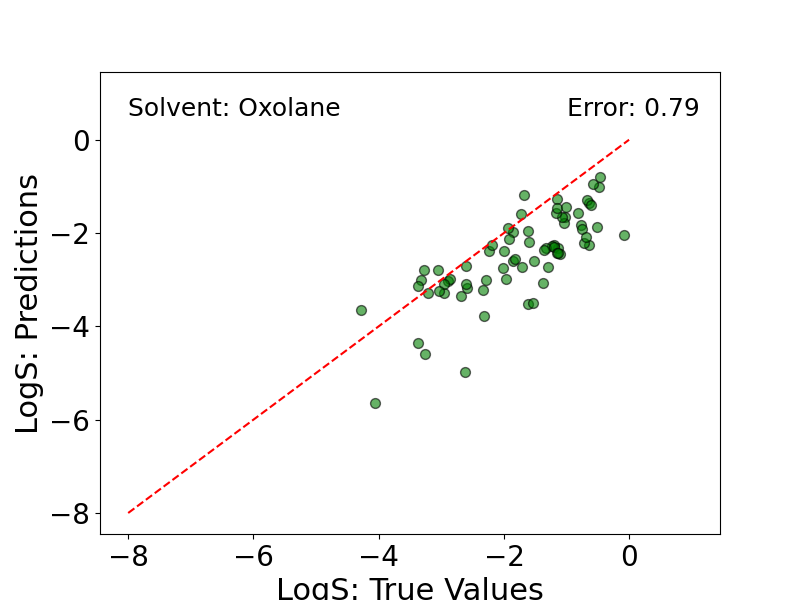}
    \caption{Predicted \textit{vs.} experimental LogS values of a few pharmaceutically important solvents}
    \label{fig:ex2}
\end{figure}

\begin{figure}
    \centering
    \includegraphics[width=1\linewidth]{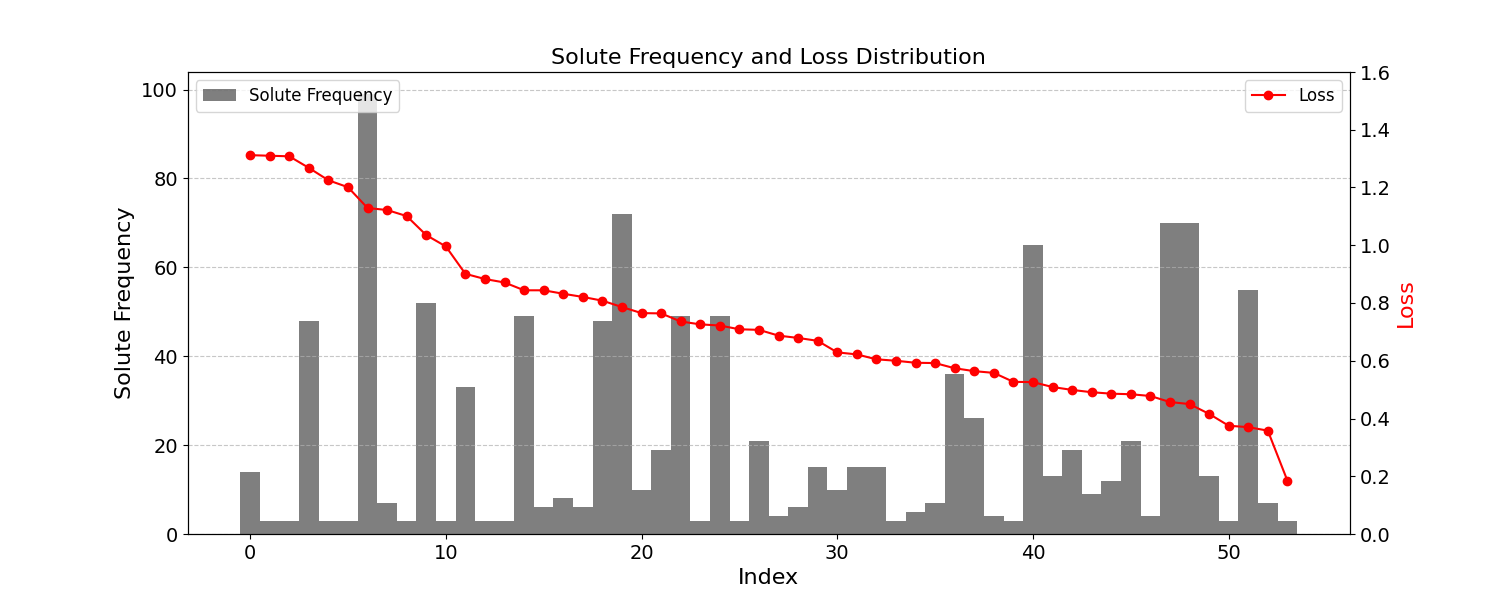}
    \caption{Solute Frequency and Loss (MAE) for each solvent by index}
    \label{fig:eval2}
\end{figure}

The GNN-based solubility prediction model demonstrated robust performance in the prediction of molecular solubility. (Figure \ref{fig:regres})  The key performance metrics, including mean absolute error (MAE), $R^2$ Score, and other relevant evaluation metrics are provided in Figures \ref{fig:regres}, \ref{fig:ex2} and \ref{fig:eval2}, respectively. The model's accuracy is evident from its ability to closely match experimental solubility values across a diverse set of solute-solvent pairs. (Table X)
Evaluation of 55 solvents showed an average loss (MAE of each solvent) of 0.7501. 80\% solvents (44 solvents) having MAE less than 1 and 58\% solvents with MAE less than 0.75. The average $R^2$ score is 0.59, but we notice the $R^2$ value is not the best indicator of the model's performance, for fewer data points in each solvent as it is highly sensitive to outliers and can be significantly affected by a few data points with extreme values, leading to an inaccurate assessment of the model's performance. 

Finally, in Table 7, we provided the solubility of a few popular drug molecules predicted using our model and compared them with the experimental values. The predicted solubility values are close to the experimental values manifesting the effectiveness of our model. 

\newpage
\begin{table}
    \centering
    \begin{tabular}{|c|c|c|c|c|}
    \hline
       Index & Molecule Name & Solvent & Predicted LogS & Expt. LogS  \\
       \hline
        1) & Aspirin & Water &-2.806&-2.699 \\	
        6) & Aspirin & Acetone &-0.101&0.025 \\
        2) & Paracetamol & Water&-2.052&-2.179  \\
        8) & Paracetamol & Methanol&-1.392&-1.387 \\
        3) & Amlodipine	& Water &-5.486&-5.611 \\
        4) & Alprazolam	& Water &-5.958&-5.489 \\
        5) & Metoprolol	&Water&-4.981&-4.535 \\	
        7) & Metformin & Water  &-1.503 &-1.744 \\
        9) & Diazepam	& Methanol &-2.406&-2.699 \\
        \hline
    \end{tabular}

    \caption{Solubility values - predicted and experimental for some common drugs}
    \label{tab:my_label}
\end{table}

\section{CONCLUSIONS}
We proposed a new method of predicting solubility that explicitly utilizes structural information of solutes and solvents and their interactions. The MolMerger algorithm provides a unique way of incorporating solute-solvent intermolecular interactions by calculating the Gasteiger charges and finding the most polar sites in the solute and solvent molecules. This method does neither need any input data from expensive quantum chemical calculations nor data from experiments. The presented method is not limited to predicting the solubility of a solute in a specific or a small set of solvents, rather it covers a large range of solute-solvent combinations predicting their solubility with an accuracy range $\sim$50-80 \%. The presented method can be improved further by incorporating more meaningful solvent interactions and temperature. Furthermore, molecular dynamics simulations-based methods\cite{liu2016using,li2018computational,bjelobrk2021solubility,bjelobrk2022solubility,reinhardt2023streamlined} can also be augmented with the proposed prediction model to calculate realistic solubility of ionic and molecular crystals.

\section{ASSOCIATED CONTENT}
\noindent
\textbf{Data Availability Statement}\\
The codes and datasets can be found in a public GitHub repository (https://github.com/).\\


\section{AUTHOR INFORMATION}
\noindent
\textbf{Corresponding Author:}\\ \textbf{Tarak Karmakar}\\ Department of Chemistry, Indian Institute of Technology, \\ Delhi 110016, New Delhi, India;\\
orcid.org/0000-0002-8721-6247;\\ Phone:+91 11 26548549; Email: tkarmakar@chemistry.iitd.ac.in\\

\noindent
\textbf{Authors}\\
\textbf{Vansh Ramani}\\Department of Chemical Engineering, Indian Institute of Technology, \\ Delhi 110016, New Delhi, India;\\

\noindent
{\bf Notes:}\\ 
The authors declare no competing financial interest.

\section{ACKNOWLEDGEMENTS}
T.K. acknowledges the Science and Engineering Research Board (SERB), New Delhi, India for the Start-up Research Grant (File No. SRG/2022/000969). We also acknowledge IIT Delhi for the Seed Grant. 

\bibliography{ref}{}
\end{document}